\documentclass{article}
\usepackage{spconf,amsmath,graphicx,hyperref}
\usepackage{lipsum}
\usepackage{booktabs}
\usepackage{threeparttable}
\usepackage{amssymb}

\newcommand{\subalign}[1]{%
  \vcenter{%
    \Let@ \restore@math@cr \default@tag
    \baselineskip\fontdimen10 \scriptfont\tw@
    \advance\baselineskip\fontdimen12 \scriptfont\tw@
    \lineskip\thr@@\fontdimen8 \scriptfont\thr@@
    \lineskiplimit\lineskip
    \ialign{\hfil$\m@th\scriptstyle##$&$\m@th\scriptstyle{}##$\hfil\crcr
      #1\crcr
    }%
  }%
}

\usepackage[utf8]{inputenc}

\title{Automatic Music Mixing using a Generative Model of Effect Embeddings}

\name{
\begin{tabular}[t]{c}
    Eloi Moliner$^{*1}$\thanks{* Work done during an internship at Sony AI. } \qquad
    Marco A. Martínez-Ramírez$^{2}$ \qquad
    Junghyun Koo$^{2}$ \qquad
    Wei-Hsiang Liao$^{2}$ \\
    \textit{Kin Wai Cheuk}$^{2}$ \qquad
    \textit{Joan Serrà}$^{2}$ \qquad 
    \textit{Vesa Välimäki}$^{1}$ \qquad
    \textit{Yuki Mitsufuji}$^{2,3}$
\end{tabular}
}
  
  \address{$^{1}$ Acoustics Lab, DICE, Aalto University, Espoo, Finland\\
  $^{2}$ Sony AI \quad $^{3}$ Sony Group Corporation}
\begin{document}
\ninept
\maketitle
\begin{abstract}
Music mixing involves combining individual tracks into a cohesive mixture, a task characterized by subjectivity where multiple valid solutions exist for the same input. Existing automatic mixing systems treat this task as a deterministic regression problem, thus ignoring this multiplicity of solutions. Here we introduce MEGAMI (Multitrack Embedding Generative Auto MIxing), a generative framework that models the conditional distribution of professional mixes given unprocessed tracks. 
MEGAMI uses a track-agnostic effects processor conditioned on per-track generated embeddings, handles arbitrary unlabeled tracks through a permutation-equivariant architecture, and enables training on both dry and wet recordings via domain adaptation. Our objective evaluation using distributional metrics shows consistent improvements over existing methods, while listening tests indicate performances approaching human-level quality across diverse musical genres.
\end{abstract}

\section{Introduction}

Music mixing combines individual instrument tracks into a final song using audio effects such as equalization, compression, panning, and reverb. This process uses both technical and subjective criteria to create a cohesive final mixture, and is crucial in modern music production~\cite{IMPbook19}. Due to its complexity, recent years have witnessed increased research in intelligent music production systems that automate or assist diverse audio engineering tasks~\cite{steinmetz2022automix, tenyearsautomix}.

Early approaches to automatic music mixing used rule-based systems~\cite{tenyearsautomix} and classical machine learning techniques~\cite{moffat19approaches}, but lacked flexibility and suitable datasets pairing dry (non-processed) recordings and wet (processed) tracks with known effect parameters. 
More recent deep learning approaches essentially fall into two main categories: direct transformation systems that predict the final mix end-to-end~\cite{martinez2021deep, martinez2022automatic}, and parameter estimation systems that predict parameters for traditional audio effects~\cite{steinmetz2020mixing}. 
The former provide powerful nonlinear mappings, but can introduce artifacts and often lack interpretability, while the latter offer controllability, but are limited to fixed effect chains with  
differentiable implementations.

A critical limitation of previous works is treating automatic mixing as a deterministic, one-to-one mapping problem. In reality, multiple valid mixes exist for any input tracks, reflecting different artistic choices and production styles~\cite{vanka2023adoption}. This one-to-many mapping is inadequately captured by regression models that learn to predict a single ``average'' mix, often resulting in overly conservative outputs suffering from regression-to-the-mean effects. Style transfer approaches~\cite{koo2023music,vanka2024diff} use reference mixtures to enable user control, but remain limited by fixed effect chains and deterministic mappings.
Additionally, replicating reference audio effects does not always equal to mixing style transfer, as mixing style encompasses artistic decisions beyond effect parameter matching.

\begin{figure}
    \centering
    \includegraphics[width=\columnwidth]{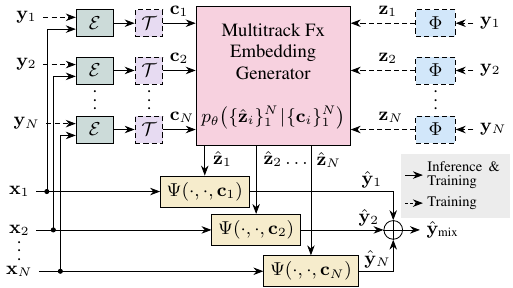}
    \caption{
Diagram of the proposed MEGAMI system. 
}

\label{fig:diagram}
\end{figure}

This paper presents MEGAMI (Multitrack Embedding Generative Auto MIxing), a generative framework for automatic mixing that operates in an effect embedding space rather than directly in the audio domain (Fig.~\ref{fig:diagram}). Unlike previous regression-based approaches, our method uses conditional diffusion models to capture the multimodal distribution of professional mixing decisions, thus handling the one-to-many nature of the task while avoiding the undesired alterations of musical content that may arise with audio-domain generative models. In particular, our framework disentangles mixing effects from musical content through learned embeddings, and enables training on both dry and wet stems via domain adaptation in the representation space. This contrasts with previous works using dry-only \cite{steinmetz2020mixing} or wet-only \cite{martinez2022automatic} stems, or signal processing-based effect normalization \cite{martinez2022automatic}.

The contributions of this work include what we believe is (1) the first generative approach to automatic mixing using conditional diffusion models. The approach features (2) a novel effect embedding factorization that separates mixing decisions from musical content, (3) a domain adaptation strategy achieved via effects removal in the CLAP embedding space (enabling more accurate training on large-scale wet-only datasets), and (4) a permutation-equivariant transformer architecture handling arbitrary numbers of unlabeled tracks. Finally, (5) we report objective improvements over existing methods and a subjective evaluation that shows the performance of MEGAMI approaching human-level quality. Contributions 1--4 are detailed in Sec.~\ref{method}, while 5 can be found in Sec.~\ref{experiments}. Sec.~\ref{conclusion} concludes the paper.

\section{MEGAMI}
\label{method}

Our goal is to approximate the conditional distribution of professional mixes $\mathbf{y}_\text{mix}$ given a set of $N$ unprocessed tracks $\mathcal{X}=\{\mathbf{x}_i\}_{i=1}^N$, that is $p(\mathbf{y}_\text{mix}\mid \mathcal{X}) $.
The MEGAMI system produces a corresponding set of processed output tracks $\hat{\mathcal{Y}}=\{ \hat{\mathbf{y}}_i \}_{i=1}^N $.
Assuming no master-bus effects, the mix is then given by $\hat{\mathbf{y}}_\text{mix} = \sum_{i=1}^N \hat{\mathbf{y}}_i$. Consequently, the distribution of the mix is fully determined by the joint distribution $p(\hat{\mathcal{Y}} \mid \mathcal{X})$. 
This distribution is potentially complex, reflecting diverse professional mixing practices. Modeling it directly is suboptimal, as mixing decisions and audio effects are entangled with musical content, which should remain unchanged.
To disentangle these factors, we introduce a latent-variable formulation in which audio effect information is explicitly separated from musical content.

\subsection{Multitrack Effect Embedding Generator} \label{sec:diff}

The core of MEGAMI is a conditional diffusion model that captures the joint distribution of a set of latent effect embeddings for all tracks in a song: 
 $p_\theta(\hat{\mathcal{Z}} \mid \mathcal{E}(\mathcal{X})) \approx p(\hat{\mathcal{Z}} \mid \mathcal{X})$.
The model generates a set of effect embeddings $\hat{\mathcal{Z}} = \{\hat{\mathbf{z}}_i\}_{i=1}^N$, $\hat{\mathbf{z}}_i \in \mathbb{R}^D$, conditioned on a corresponding set of features $\mathcal{C}=\{\mathbf{c}_i \}_{i=1}^N$, where the embeddings $\mathbf{c}_i \in \mathbb{R}^C$ are extracted from the input tracks through an audio encoder: $\mathbf{c}_i = \mathcal{E}(\mathbf{x}_i)$ (Fig.~\ref{fig:diagram}).
In our implementation, $\mathcal{E}$ corresponds to the CLAP encoder~\cite{wu2023large}, 
which allows the diffusion model to infer track semantics, such as musical instrument type, without requiring explicit labels. 
At training time,  we use embeddings $\mathcal{Z} = \{\mathbf{z}_i\}_{i=1}^N$ as ground truth, which are obtained by $ \mathbf{z}_i= \Phi(\mathbf{y}_i)$, where $\mathbf{y}_i$ is a wet track from a dataset, and $\Phi$ is a feature extractor designed to be injective with respect to production style, so that $\mathbf{z}_i$ captures the unique mixing characteristics of track $i$.
For this purpose, we use FxEncoder++~\cite{yeh2025fx}, a model trained via contrastive learning to disentangle effect characteristics from musical content.

Both embeddings $\mathcal{C}$ and $\hat{\mathcal{Z}}$ capture time-invariant, track-level characteristics rather than frame-level features, reflecting the assumption that the applied effects remain consistent over the duration of a track.
%The 2048-dimensional vector is taken prior to the final layer.
To better capture dynamics and gain-related information, each embedding in $\hat{\mathcal{Z}}$ is augmented with a set of dynamic and stereo features, including log-RMS, crest factor, dynamic spread, stereo width, and stereo imbalance. 
These features are standardized and
expanded into a 64-dimensional vector using a deterministic Fourier feature–based transformation \cite{tancik2020fourier}.
%implemented with a DCT-style cosine basis.
This transformation is approximately invertible, allowing the original features to be recovered from the embedding.
The resulting vector is then scaled and concatenated to the 2048-dimensional embedding from FxEncoder++.

Following the EDM formulation \cite{karras2022elucidating}, we define a forward diffusion process over the set of effect embeddings $\hat{\mathcal{Z}}$, independently corrupting each element with Gaussian noise:
\begin{equation*}
\hat{\mathbf{z}}_{i,\tau} = \hat{\mathbf{z}}_i + \sigma(\tau)  \boldsymbol{\epsilon}_i, \quad \boldsymbol{\epsilon}_i \sim \mathcal{N}(\mathbf{0}, \mathbf{I}_D), \quad i=1,\dots,N,
\end{equation*}
where $\sigma(\tau)=\tau$,  $\tau \in [0,T]$, $\hat{\mathbf{z}}_{i,0} = \hat{\mathbf{z}}_i$, and $\hat{\mathbf{z}}_{i,T} \sim \mathcal{N}(\hat{\mathbf{z}}_i, \sigma^2(T) \mathbf{I}_D)$. 
Assuming that $\mathcal{N}(\hat{\mathbf{z}}_i, \sigma^2(T) \mathbf{I}_D) \approx  \mathcal{N}(\mathbf{0}, \sigma^2(T) \mathbf{I}_D)$, sampling can be initialized by drawing $\hat{\mathcal{Z}}_T \sim \mathcal{N}(\mathbf{0}, \sigma^2(T) \mathbf{I}_D)$. From this initialization, the reverse diffusion process is applied to progressively refine the samples, 
generating diverse and coherent sets of embeddings $\hat{\mathcal{Z}}$.
The reverse process is parameterized via the probability flow ODE \cite{karras2022elucidating,songscore}:
\begin{equation*}\label{eq:pfode}
d\hat{\mathcal{Z}}_\tau = -\tau \, \nabla_{\hat{\mathcal{Z}}_\tau} \log p_\theta(\hat{\mathcal{Z}}_\tau \mid \mathcal{E}(\mathcal{X})) \, d\tau, \quad \hat{\mathcal{Z}}_\tau = \{\hat{\mathbf{z}}_{i,\tau}\}_{i=1}^N,
\end{equation*}
where the score function is approximated by a neural network $s_\theta(\hat{\mathcal{Z}}_\tau, \mathcal{C}, \tau) \approx \nabla_{\hat{\mathcal{Z}}_\tau} \log p_\theta(\hat{\mathcal{Z}}_\tau \mid \mathcal{C})$ trained via denoising score matching \cite{vincent2011connection}, following the standard choices in \cite{karras2022elucidating}.

The score model $s_\theta(\hat{\mathcal{Z}}_\tau, \mathcal{C}, \tau)$ is designed to be permutation-equivariant, so that reordering $\mathcal{C}$ produces a corresponding reordering of the output embeddings. We implement this using a transformer that performs self-attention across the elements of $\hat{\mathcal{Z}}_\tau$ and cross-attention with $\mathcal{C}$.
To preserve track-to-track correspondence, we concatenate a one-hot vector representing the track position $i$ to both the $i$-th element of $\hat{\mathcal{Z}}_\tau$ and the $i$-th element of $\mathcal{C}$. During training, the order of tracks is randomly permuted, which prevents the model from associating a fixed sequence position with a specific semantic role.
To accommodate sets of varying size (i.e., songs with a different number of tracks), both $\hat{\mathcal{Z}}_\tau$ and $\mathcal{C}$ are padded with zeros up to the fixed maximum size $N=14$ during training, and attention/cross-attention masks are applied to ensure that the transformer ignores the padded elements.

\subsection{Training with Wet-Only Tracks via Domain Adaptation}
\label{sec:da}

The score model $s_\theta$ ideally requires paired dry and wet multitracks for training. Such datasets are extremely limited, whereas professionally processed, stem-separated music—where each track is isolated and summing all stems forms a complete mix—is more accessible. Public datasets exist for source separation, and internally we maintain a large collection of stem-separated recordings. 
Previous work~\cite{martinez2022automatic} proposed signal processing-based effect normalization techniques to repurpose wet multitrack recordings for the training of automatic music mixing systems. However, such methods cannot always fully remove effect variance due to inherent signal processing limitations and effect interdependencies, thus limiting generalization when training exclusively on processed stems.
Instead, we propose a domain adaptation strategy in the content representation space.

While the content encoder $\mathcal{E}$, based on CLAP, is primarily trained for audio–text alignment, it can retain traces of production effects, which are undesirable here. If not removed, this information leaks into the conditioning, causing the model to fail at learning effect embeddings. To suppress this, we introduce a domain adaptor $\mathcal{T}$ that performs effect removal by mapping wet embeddings toward the dry domain (Fig.~\ref{fig:diagram}). 
Formally, let $p_x$ denote the distribution of dry tracks and $p_y$ the distribution of processed tracks. We introduce a deterministic domain adaptor $\mathcal{T}$ that aligns the two embedding distributions. Specifically, we require that the distribution of adapted wet embeddings $\mathcal{T}(\mathcal{E}(\mathbf{y}))$ for $\mathbf{y} \sim p_y$, after smoothing with a Gaussian kernel $\varphi_{\sigma_{\mathcal{T}}} = \mathcal{N}(\mathbf{0}, \sigma_\mathcal{T}^2 \mathbf{I})$, approximately matches the distribution of dry embeddings $\mathcal{E}(\mathbf{x})$ for $\mathbf{x} \sim p_x$:
\begin{equation*}
\varphi_{\sigma_\mathcal{T}}
\ast
\int \delta(\mathbf{c}-\mathcal{T}(\mathcal{E}(\mathbf{y})))p_y(\mathbf{y}) d\mathbf{y}
\approx
\varphi_{\sigma_\mathcal{T}}
\ast
\int \delta(\mathbf{c}-\mathcal{E}(\mathbf{x}))p_x(\mathbf{x}) d\mathbf{x},
\end{equation*}
where $\ast$ is the convolution operator and $\delta(\cdot)$ is a Dirac delta distribution.

The adaptor $\mathcal{T}$ is implemented as a multi-layer perceptron with two layers, and is trained on a small dataset of single-track dry/wet pairs $(\mathbf{x}, \mathbf{y})$ by minimizing the $L_2$ loss between 
the adapted wet representation $\mathcal{T}(\mathcal{E}(\mathbf{y}))$
and the dry representation $\mathcal{E}(\mathbf{x})$.
The Gaussian kernel $\varphi_{\sigma_\mathcal{T}}$ reduces domain mismatch by smoothing the distributions, and additionally acts as a data augmentation strategy that reduces overfitting when training $p_\theta$. It is implemented by adding Gaussian noise to the embeddings $\mathbf{c}$ at inference time.
With the adaptor in place, the diffusion model can be trained directly on processed wet stems without requiring paired dry multitracks.

\subsection{Effect Processor}
Each processed track $\hat{\mathbf{y}}_i$ is reconstructed independently via a deterministic effect processor $\hat{\mathbf{y}}_i = \Psi(\mathbf{x}_i, \hat{\mathbf{z}}_i, \mathbf{c}_i)$, which applies the effects encoded in $\hat{\mathbf{z}}_i$ to the input track $\mathbf{x}_i$ (Fig.~\ref{fig:diagram}). 
 The processor is also conditioned on the content embedding $\mathbf{c}_i= \mathcal{E}(\mathbf{x}_i) $ extracted from the dry input. The purpose of conditioning the model with content embeddings $\mathbf{c}_i$ (CLAP) is to improve adaptability across diverse musical content (e.g., drums, vocals, etc.).

We employ a single, track-agnostic black-box predictive model for $\Psi$, as in \cite{koo2023music}.
This implies that the per-track conditional distribution is approximated as $p(\hat{\mathbf{y}}_i \mid \mathbf{x}_i, \hat{\mathbf{z}}_i) \approx \delta(\hat{\mathbf{y}}_i - \Psi(\mathbf{x}_i, \hat{\mathbf{z}}_i, \mathbf{c}_i))$. 
The Dirac delta assumption indicates that all variability in mixing style is captured by $\hat{\mathbf{z}}_i$ and that, conditioned on $\hat{\mathbf{z}}_i$, the mapping $\mathbf{x}_i \mapsto \hat{\mathbf{y}}_i$ is effectively deterministic.
Accordingly, the conditional distribution over the set of processed tracks can be expressed as
\begin{equation*}
p_{\theta}(\hat{\mathcal{Y}} \mid \mathcal{X}) = \int \prod_{i=1}^N \delta\big(\hat{\mathbf{y}}_i - \Psi(\mathbf{x}_i, \hat{\mathbf{z}}_i, \mathbf{c}_i)\big) \, p_\theta\big(\hat{\mathcal{Z}} \mid \mathcal{E}(\mathcal{X})\big) \, d\hat{\mathcal{Z}},
\end{equation*}
where $p_\theta$ captures the variability and multimodality of professional mixes, and $\Psi$ ensures that each track is reconstructed consistently given its generated embedding. 
In practice, we implement $\Psi$ as a temporal convolutional network (TCN) as in \cite{steinmetz2020mixing}.
Stereo inputs are first converted to mono, since stereo information such as panning and stereo width is considered part of the mixing style and thus encoded in $\hat{\mathbf{z}}_i$. Consequently, $\Phi$ is responsible for upmixing the mono inputs to stereo.
Next, the input tracks are EQ normalized using average spectral features following~\cite{martinez2022automatic}. The inputs are RMS normalized before and after EQ. 
The latent effect embedding $\hat{\mathbf{z}}_i$ and the content embedding $\mathbf{c}_i$ are concatenated and injected into the network via feature-wise linear modulation.

During training, the target signals $\mathbf{y}_i$ are also RMS-normalized to stabilize optimization. However, the corresponding training embeddings $\mathbf{z}_i = \Phi(\mathbf{y}_i)$ are extracted prior to normalization.
At inference, the track-level channel-wise RMS is retrieved from the embedding and used to apply per-channel gain to the audio signals.
RMS values are computed over overlapping windows, and the maximum across windows is stored as the track-level gain reference. 
Following Koo et al.~\cite{koo2023music}, we use the multi-scale spectral loss for training.
In addition, we include a secondary deep feature loss, computed as the cosine distance between the FxEncoder++
embeddings of the output $\Phi(\hat{\mathbf{y}})$ and the ground truth signal $\Phi(\mathbf{y})$.
This enforces the reconstructed track to reflect the effect characteristics encoded in $\mathbf{z}_i$.

\section{Experiments and Results}
\label{experiments}

\begin{figure*}[t]
    \centering
    \includegraphics[trim=7 11 6 7 , clip, width=0.95\textwidth]{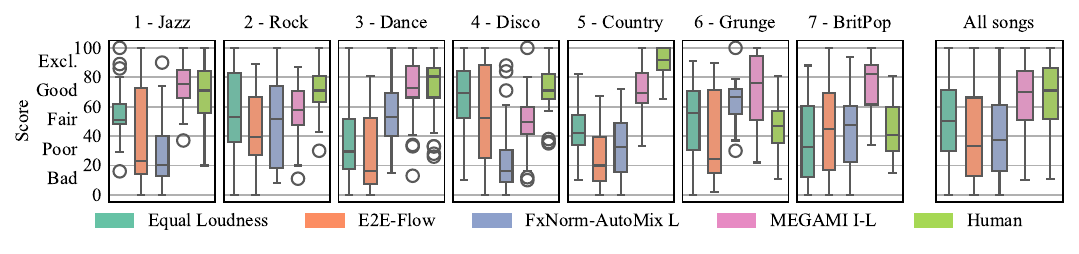}
    \vspace{-3pt}
    \caption{
    Boxplots of subjective listening test scores for each song individually and for all songs combined, showing that MEGAMI approaches the quality of a human mixing engineer and, in some cases, exceeds it.}
    
    \label{fig:boxplots}
\end{figure*}

\subsection{Datasets}\label{sec:data}

We curate several datasets to train the different components and configurations of our system:
\begin{itemize}
   \setlength\leftskip{-1.2em}
    \item \textit{Internal-Small (I-S)}:
    Internal dataset of about 400 professionally mixed songs with available pairs of isolated dry and wet stems. Each song contains up to $N=14$ separated tracks.
    \item \textit{Internal-Large (I-L)}:
    Internal dataset of about 20\,k professionally mixed songs with wet stems only.
    This dataset contains the songs in \textit{I-S}, and mainly consists of Western popular music.
    \item \textit{Public (P)}:
    Collection of openly licensed
    multitracks from MoisesDB \cite{pereira2023moisesdb} and MedleyDB \cite{bittner2014medleydb}, totaling 248 multitracks with wet stems in the training split.
     Since the number of stems is inconsistent across songs, we randonly regroup instrumentally related stems as needed, enforcing an upper limit of $N=14$ stems per song. 
    This also serves as data augmentation by exposing the model to different stem groupings. 
    \item \textit{Public-Dry (P-D)}: 
    Dataset built of dry or semi-dry single tracks from openly available sources. Those are MedleyDB \cite{bittner2014medleydb}, OpenSinger \cite{huang2021multi}, IDMT-SMT Drums/Bass/Guitar \cite{dittmar2014real, abesser2010feature, kehling2014automatic}, GuitarSet \cite{xi2018guitarset}, and Aalto anechoic orchestra \cite{patynen2008anechoic}.
    \item \textit{Evaluation Benchmark}: Internal benchmark of 59 multitracks with available dry/wet stem pairs, none of which were used for training. From each song, we extracted 10 segments of 11.9\,s, resulting in 590 test examples used for objective evaluation.
    
\end{itemize}

\subsection{Implementation Details}

The proposed system is composed of five models. Three of them were trained independently for this work: the multitrack effect embedding score model $s_\theta$, the effect processor $\Psi$, and the domain adaptor $\mathcal{T}$. Each of these models requires distinct datasets and implementation specifics.
In addition, two pretrained feature extractors are integrated into the system: FxEncoder++ \cite{yeh2025fx} ($\Phi$) and CLAP \cite{wu2023large} ($\mathcal{E}$). For CLAP, we employ a publicly released checkpoint trained specifically on music data.

We train separate instances of the score model $s_\theta$ on datasets \textit{I-S}, \textit{I-L}, and \textit{P}. For \textit{I-S}, dry tracks are used directly, while for \textit{I-L} and \textit{P}, training is performed on wet stems using the domain adaptation strategy described in Sec.~\ref{sec:da}. 
The effect processor $\Phi$ and domain adaptor $\mathcal{T}$ require paired dry/wet data. For this purpose, we use \textit{I-S}, augmenting the wet stems with random effect chains comprising equalization, compression, reverberation, and Haas delay \cite{zolzerdafx}. 
For the public version of the system, $\Phi$ and $\mathcal{T}$ are trained on \textit{P-D}, where wet counterparts are generated synthetically with the same random effect chains, combined with publicly available impulse responses from ReverbFx~\cite{richter2025reverbfx}, ASH~\cite{pearce_2019_2638644}, and Arni~\cite{prawda2022robust}.

All models are trained and evaluated on stereo signals at a sampling rate of 44.1\,kHz, using segments of 11.9\,s (although they can process longer audio at inference).
Inputs to $\Phi$ and $\mathcal{E}$, however, are converted to mono by averaging the two channels, while the targets remain stereo. 
The score model $s_\theta$ is implemented as a transformer architecture with approximately 70\,M~parameters, 
while the effect processor $\Psi$ employs a TCN architecture with roughly 9\,M~parameters.
Further details will be provided in the code repository\footnote{\href{https://github.com/SonyResearch/MEGAMI}{https://github.com/SonyResearch/MEGAMI}}.

\subsection{Baselines} \label{sec:baselines}

We compare our method against several baselines. As a simple reference, we include \textit{Equal Loudness}, which balances tracks using perceptual loudness. We also evaluate \textit{Fxnorm-Automix}~\cite{martinez2022automatic} with both the publicly available ``S'' version and the larger internal ``L'' version.
%using its publicly released pretrained weights; both the publicly available S-version and the larger internal L-version are considered. 
In addition, we include \textit{DMC} \cite{steinmetz2020mixing}, which predicts only gain and panning operations, retrained on the \textit{I-S} dataset, as well as \textit{MixWaveUnet} \cite{martinez2021deep}, a black-box predictive model also retrained on \textit{I-S}.

We further introduce an end-to-end baseline, \textit{E2E-Flow}, inspired by \cite{wu2024diffusion}. Unlike our factorized approach, this model directly approximates the conditional distribution $p(\hat{\mathcal{Y}} \mid \mathcal{X})$. It learns a flow that maps stacked, RMS-normalized dry stems (grouped into four stems and augmented with Gaussian noise) to the distribution of wet stems. Training is performed via flow matching \cite{lipmanflow} in the complex STFT domain, with stereo tracks stacked along the channel dimension in a fixed order. We retrain \textit{E2E-Flow} on the \textit{I-S} dataset.
In addition, we consider two Oracle baselines informed by track-wise embeddings extracted from ground-truth human-mixed stems. These baselines isolate the performance of the effect processor and are evaluated with models trained on both the \textit{I-S} and \textit{P-D} datasets.

\subsection{Objective Evaluation}

Conventional pairwise metrics, which compare system outputs to a single human-produced reference, are not well suited for evaluating generative approaches. Instead, we use a distributional distance between the set of human mixes in the benchmark and the set of mixes produced by each system. We employ the Kernel Audio Distance (KAD), a distributional metric based on Maximum Mean Discrepancy between embeddings. All embeddings are extracted from the final audio mixture, ensuring a fair comparison across systems with different input configurations.

We compute KAD using several embedding representations: AFxRep \cite{steinmetz2024st} and FxEncoder \cite{koo2023music}, both designed for effect-related analysis, as well as FxEncoder++ \cite{yeh2025fx} and CLAP embeddings. The latter two are components of MEGAMI itself and are therefore reported as potentially biased. The results presented in Table~\ref{table:objective} show that MEGAMI consistently outperforms all automatic mixing baselines, achieving the lowest distributional distances to human mixes. Among data-dependent variants, the \textit{I-L} system outperforms \textit{I-S} and \textit{P} (except on CLAP), highlighting both the advantage of training with larger datasets and the effectiveness of the domain adaptation strategy.
We also see that, while MEGAMI does not fully match the performance of its oracle variant, it comes close to it.

\subsection{Subjective Evaluation}

The listening test was conducted in isolated booths at the Aalto Acoustics Lab using a multi-stimulus design. Participants rated the production quality of several mixes, with the input dry tracks (grouped into four stems) presented as a reference.
Seven songs from different genres were included, each presented twice, resulting in a total of 14 pages. To keep the task manageable, the number of stimuli to be evaluated was limited to five per page: a human reference mix, the Equal Loudness baseline, E2E-Flow, FxNorm-AutoMix L, and the proposed MEGAMI I-L. Participants were encouraged, though not forced, to use the full rating scale.
Twelve volunteers participated in the test, six of whom reported experience in audio production, mixing, or mastering. The test lasted approximately 30\,min on average, and was generally not considered fatiguing.

\begin{table}[t]
\vspace{-6pt}
\caption{Kernel Audio Distance (KAD) computed with different embeddings. The best results are shown in bold (lower is better). The asterisk (*) denotes pretrained weights from the original work.
}
\label{table:objective}
\vspace{2mm}
\resizebox{\columnwidth}{!}{%
\begin{threeparttable}
\begin{tabular}{@{}lcc|cccc@{}}
\toprule
                        & &  & \multicolumn{4}{c}{\textbf{KAD across embeddings}}      \\ 
\textbf{Method} & Data                         & $N$ & AFxRep & FxEnc & FxEnc++ & CLAP \\ \midrule
Equal Loudness &-          & -     & 38.08  & 49.31     & 35.74       & 4.96 \\ 
FxNorm-AutoMix S & *     & 4           & 14.22  & 6.00      & 18.37       & 2.38 \\
FxNorm-AutoMix L &*     & 4           & 11.77  & 2.64      & 8.02        & 1.31 \\ 
MixWaveUNet  & \textit{I-S} & 4           & 12.99  & 57.96     & 23.45       & 1.76 \\
DMC &\textit{I-S}       & 14     & 9.93   & 75.74     & 36.7        & 3.16 \\
E2E-Flow & \textit{I-S}      & 4           & 17.15   & 5.44      & 14.98       & 5.48 \\
MEGAMI &\textit{I-S}   & 14 & 5.89 & 1.86 & 7.44 & \textbf{0.38} \\ 
MEGAMI &\textit{I-L}  & 14 & \textbf{5.21} & \textbf{1.72} & \textbf{3.90} & 0.84 \\
MEGAMI &\textit{P} & 14 & 7.32 & 3.28 &  9.85 & 1.12 \\ \midrule    
MEGAMI Oracle & \textit{I-S} & - & 4.61 & 1.51 & 2.34 & 0.42\\
MEGAMI Oracle & \textit{P-D} & - &  5.69 & 0.94 & 3.35 & 0.91 \\ \bottomrule
\end{tabular}

\end{threeparttable}}

\end{table}

The results, shown in Fig.~\ref{fig:boxplots}, are reported both per song (left) and aggregated across all examples (right). Given the subjective nature of the task, the confidence intervals are wide and the results are strongly song-dependent. Participants penalized acoustic artifacts in the FxNorm-AutoMix and E2E-Flow baselines, which led to low scores for certain songs. Interestingly, the Equal Loudness baseline performed unexpectedly well on some examples, achieving comparatively high scores (e.g., the Disco track). In several cases, the human reference mix received lower ratings than MEGAMI, which was perceived as clearly superior (e.g., Grunge and BritPop). Overall, the proposed MEGAMI system approaches the quality of human mixes and outperforms the compared baselines in the majority of cases. We encourage the reader to listen to the public sound examples\footnote{\href{https://SonyResearch.github.io/MEGAMI}{https://SonyResearch.github.io/MEGAMI}}, which correspond to the material used in the listening test.

\section{Conclusion}
\label{conclusion}

We present MEGAMI, the first generative framework for automatic music mixing.
MEGAMI addresses key limitations of existing systems by incorporating a conditional diffusion model operating in an effect embedding space to capture the multiplicity of mixing decisions, domain adaptation to enable training on wet-only datasets, and a permutation-equivariant architecture to handle an arbitrary number of unlabeled tracks.
The conducted evaluations presented in this paper demonstrate consistent improvements over existing methods in both objective and subjective metrics.

A novel generative framework opens many potential opportunities for future work. 
Since performance scales with dataset size, our method can be used to generate synthetic mix datasets from available dry or wet multitrack recordings.
The TCN effect processor could be replaced by parameter estimation networks, improving interpretability.
Time-varying embeddings could enable dynamic mixing decisions.
Album-level coherence could be addressed by modeling consistent mixing characteristics across multiple songs.

\section{Acknowledgements}
We acknowledge the computational resources from Aalto Science-IT program. We thank the participants of the listening test.

\section{Compliance with Ethical Standards}

The listening test with volunteers, involving only anonymous and non-sensitive audio ratings, was conducted in accordance with the Declaration of Helsinki (1975, revised 2000).

\bibliographystyle{IEEEbib}
\bibliography{strings,refs}

@string{icassp = "Proc. ICASSP"}

@string{ismir = "Proc. ISMIR"}

@string{iclr = "Proc. ICLR"}

@string{nips = "Proc. NeurIPS"}

@inproceedings{yeh2025fx,
  title={{Fx-Encoder+}+: Extracting Instrument-Wise Audio Effects Representations from Mixtures},
  author={Y.-T. Yeh and J. Koo and M. A. Martínez-Ramírez and W.-H. Liao and Y.-H. Yang and Y. Mitsufuji},
  booktitle=ismir,
  year={2025}
}

@inproceedings{wu2024diffusion,
  title={Diffusion Models for Automatic Music Mixing},
  author={Wu, X. and Horner, A.},
  booktitle={Proc. IEEE Int. Conf. Big Data (BigData)},
  pages={3242--3247},
  year={2024},
}

@inproceedings{martinez2022automatic,
  title={Automatic music mixing with deep learning and out-of-domain data},
  author={M. A. Martínez-Ramírez and W.-H. Liao and G. Fabbro and S. Uhlich and C. Nagashima and Y. Mitsufuji},
  booktitle={Proc. ISMIR},
  year={2022}
}

@inproceedings{wu2023large,
  title={Large-scale contrastive language-audio pretraining with feature fusion and keyword-to-caption augmentation},
  author={Y. Wu and K. Chen and T. Zhang and Y. Hui and T. Berg-Kirkpatrick and S. Dubnov},
  booktitle=icassp,
  pages={1--5},
  year={2023},
}

@inproceedings{karras2022elucidating,
  title={Elucidating the design space of diffusion-based generative models},
  author={Karras, T. and Aittala, M. and Aila, T. and Laine, S.},
  booktitle=nips,
  volume={35},
  pages={26565--26577},
  year={2022}
}

@article{vincent2011connection,
  title={A connection between score matching and denoising autoencoders},
  author={Vincent, P.},
  journal={Neural Computation},
  volume={23},
  number={7},
  pages={1661--1674},
  year={2011},
  publisher={MIT Press}
}

@inproceedings{songscore,
  title={Score-Based Generative Modeling through Stochastic Differential Equations},
  author={Y. Song and J. Sohl-Dickstein and D. P. Kingma and A. Kumar and S. Ermon and B. Poole},
  booktitle=iclr,
  pages =  {37799-37812},
  year={2023}
}

@inproceedings{pereira2023moisesdb,
  title={{MoisesDB}: A Dataset for Source Separation Beyond 4 Stems},
  author={I. G. Pereira and F. Araujo and F. Korzeniowski and R. Vogl},
  booktitle={Proc. ISMIR},
  year={2023}
}

@inproceedings{bittner2014medleydb,
  title={{MedleyDB}: A multitrack dataset for annotation-intensive {MIR} research},
  author={R. M. Bittner and J. Salamon and M. Tierney and M. Mauch and C. Cannam and J. P. Bello},
  booktitle={Proc. ISMIR},
  volume={14},
  pages={155--160},
  year={2014}
}

@inproceedings{huang2021multi,
  title={Multi-singer: Fast multi-singer singing voice vocoder with a large-scale corpus},
  author={Huang, R. and Chen, F. and Ren, Y. and Liu, J. and Cui, C. and Zhao, Z.},
  booktitle={Proc. ACM Int. Conf. Multimedia},
  pages={3945--3954},
  year={2021}
}

@inproceedings{dittmar2014real,
  title={Real-Time Transcription and Separation of Drum Recordings Based on {NMF} Decomposition},
  author={Dittmar, C. and G{\"a}rtner, D.},
  booktitle={Proc. DAFx},
  pages={187--194},
  year={2014}
}

@inproceedings{abesser2010feature,
  title={Feature-based extraction of plucking and expression styles of the electric bass guitar},
  author={Abe{\ss}er, J. and Lukashevich, H. and Schuller, G.},
  booktitle=icassp,
  pages={2290--2293},
  year={2010},
}

@inproceedings{kehling2014automatic,
  title={Automatic Tablature Transcription of Electric Guitar Recordings by Estimation of Score-and Instrument-Related Parameters},
  author={Kehling, C. and Abe{\ss}er, J. and Dittmar, C. and Schuller, G.},
  booktitle={Proc. DAFx},
  pages={219--226},
  year={2014}
}

@inproceedings{xi2018guitarset,
  title={{GuitarSet}: A Dataset for Guitar Transcription},
  author={Xi, Q. and Bittner, R. M. and Pauwels, J. and Ye, X. and Bello, J. P.},
  booktitle={Proc. ISMIR},
  pages={453--460},
  year={2018}
}

@article{patynen2008anechoic,
  title={Anechoic recording system for symphony orchestra},
  author={P{\"a}tynen, J. and Pulkki, V. and Lokki, T.},
  journal={Acta Acustica united with Acustica},
  volume={94},
  number={6},
  pages={856--865},
  year={2008},
  publisher={European Acoustics Association}
}

@article{martinez2021deep,
  title={A deep learning approach to intelligent drum mixing with the {Wave-U-Net}},
  author={Martinez-Ramirez, M. A. and Stoller, D. and Moffat, D.},
  journal={J. Audio Eng. Soc.},
  volume={69},
  number={3},
  pages={142},
  year={2021},
  publisher={Audio Engineering Society}
}

@inproceedings{lipmanflow,
  title={Flow Matching for Generative Modeling},
  author={Lipman, Y. and Chen, R. T. Q. and Ben-Hamu, H. and Nickel, M. and Le, M.},
  booktitle={Proc. Int. Conf. Learning Representations},
  year={2023}
}

@inproceedings{steinmetz2024st,
  title={{ST-ITO}: Controlling audio effects for style transfer with inference-time optimization},
  author={Steinmetz, C. J. and Singh, S. and Ibnyahya, I. and Yuan, S. and Benetos, E. and Reiss, J. and others},
  year={2024},
  booktitle={Proc. ISMIR}
}

@inproceedings{vanka2024diff,
  title={Diff-{MST}: Differentiable Mixing Style Transfer},
  author={Vanka, S. and Steinmetz, C. and Rolland, J. B. and Reiss, J. and Fazekas, G. and others},
  year={2024},
  booktitle={Proc. ISMIR}
}

@book{IMPbook19,
	Author = {Stables, R. and Reiss, J. D. and De~Man, B.},
	Publisher = {Focal Press},
	Title = {Intelligent Music Production},
	Year = {2019}}

@misc{steinmetz2022automix,
    Author = {C. J. Steinmetz and S. S. Vanka and M. A. Martínez-Ramírez and G. Bromham},
    Title = {Deep Learning for Automatic Mixing},
    Year = {2022},
    howpublished = {ISMIR Tutorial. Available: \url{https://dl4am.github.io/tutorial}},
}

@inproceedings{tenyearsautomix,
	Author = {B. De Man and J. D. Reiss and R. Stables},
	Booktitle = {Proc. 3rd AES Workshop on Intelligent Music Production},
	Title = {Ten Years of Automatic Mixing},
	month = {Sep.},
	Year = {2017}}

@article{moffat19approaches,
	Author = {D. Moffat and M. B. Sandler},
	Journal = {Arts},
	Month = {Sep.},
	Number = {5},
	Pages = {14},
	Title = {Approaches in Intelligent Music Production},
	Volume = {8},
	url = {https://doi.org/10.3390/arts8040125},
	Year = {2019}}

@inproceedings{steinmetz2020mixing,
            title={Automatic multitrack mixing with a differentiable mixing console of neural audio effects},
            author={Steinmetz, C. J. and Pons, J. and Pascual, S. and Serrà, J.},
            booktitle=icassp,
            year={2021}}

@inproceedings{vanka2023adoption,
title={Adoption of {AI} Technology in the Music Mixing Workflow: An Investigation},
author={S. S. Vanka and M. Safi and J.-B. Rolland and G. Fazekas},
booktitle={Proc. AES 154th Convention},
year={2023}
}

@inproceedings{koo2023music,
title={Music Mixing Style Transfer: A Contrastive Learning Approach to Disentangle Audio Effects},
author={Koo, J. and Martinez-Ramirez, M. A and Liao, W-H. and Uhlic, S. and Lee, K. and Mitsufuji, Y.},
booktitle=icassp,
year={2023},
}

@inproceedings{richter2025reverbfx,
  title={{ReverbFX}: A Dataset of Room Impulse Responses Derived from Reverb Effect Plugins for Singing Voice Dereverberation},
  author={Richter, J. and Svajda, T. and Gerkmann, T.},
  booktitle={ITG Conference on Speech Communication},
  year={2025}
}

@article{prawda2022robust,
  title={Robust selection of clean swept-sine measurements in non-stationary noise},
  author={Prawda, K. and Schlecht, S. J. and V{\"a}lim{\"a}ki, V.},
  journal={J. Acoust. Soc. Amer.},
  volume={151},
  number={3},
  pages={2117--2126},
  year={2022},
  publisher={AIP Publishing}
}

@dataset{pearce_2019_2638644,
  author       = {Pearce, S.},
  title        = {Audio Spatialisation for Headphones -- {I}mpulse
                   Response Dataset
                  },
  month        = apr,
  year         = 2019,
  note    = {Zenodo. Available from \url{https://doi.org/10.5281/zenodo.2638644}},
  version      = {5.1.0},
  doi          = {10.5281/zenodo.2638644},
}

@article{tancik2020fourier,
  title={Fourier features let networks learn high frequency functions in low dimensional domains},
  author={M. Tancik and P. Srinivasan and B. Mildenhall and S. Fridovich-Keil and N. Raghavan and U. Singhal and R. Ramamoorthi and J. Barron and R. Ng},
  journal=nips,
  volume={33},
  pages={7537--7547},
  year={2020}
}

@article{zolzerdafx,
  title={DAFX: Digital Audio Effects},
  author={Z{\"o}lzer, Udo},
  publisher={Wiley Online Library}
}

\end{document}